\documentstyle[aps,twocolumn,floats,epsf]{revtex}

\def\P{Poincar\' e }
\def\be{\begin{equation}}
\def\ee{\end{equation}}
\def\etal{{\em et al.}}

\begin{document}
\title{Universal behaviour of entrainment due to coherent structures in turbulent
shear flow}
\author{Rama Govindarajan}
\address{Fluid Dynamics Unit, Jawaharlal Nehru Centre for Advanced Scientific
Research, Jakkur, Bangalore 560064, India.
E-mail: rama@jncasr.ac.in}

\twocolumn[\hsize\textwidth\columnwidth\hsize\csname@twocolumnfalse%
\endcsname

\maketitle

\begin{abstract}
I suggest a solution to a persistent mystery in the physics of turbulent shear 
flows: cumulus clouds rise to towering heights, practically without entraining
the ambient medium, while apparently similar turbulent jets in general lose 
their identity within a small distance through entrainment and mixing. 
From dynamical systems computations on a model chaotic vortical flow, I 
show that entrainment and mixing due to coherent structures depend
sensitively on the relative speeds of different portions of the flow. 
A small change in these speeds, effected for example by heating, drastically
alters the sizes of the KAM tori and the chaotic mixing region. The entrainment 
rate and, hence, the lifetime of a turbulent shear flow, shows a {\em universal, 
non-monotone} dependence on the heating. 
\end{abstract}
]

Turbulent shear flows \cite{pope} in nature and in the laboratory are 
anisotropic and inhomogeneous on large enough length scales \cite{inhom1,inhom2}.  
Among their most important properties is {\em entrainment} \cite{roshko}, {\it viz.},
the advection of the ambient non-turbulent medium into the body of the turbulent 
flow, and the consequent expansion of the turbulent region.
Coherent structures \cite{coh1,coh2,coh3} at the edges of the turbulent 
region are crucial to entrainment, but their role in turbulent transport 
in general is only now beginning to be understood \cite{toroc}. 
\begin{figure}
\epsfxsize=0.95\columnwidth
\centerline{\epsfbox{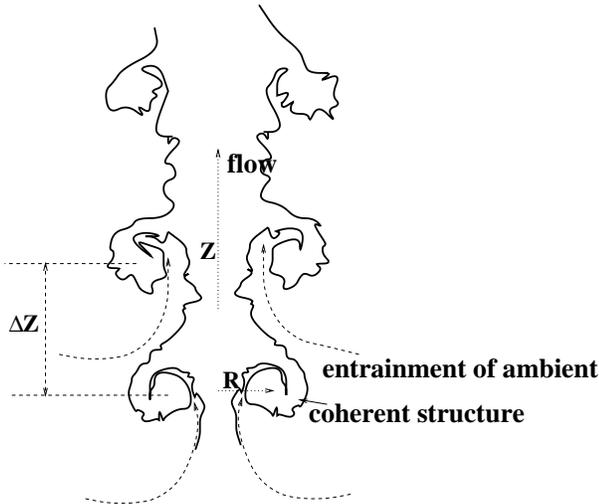}}
\caption[]{Schematic sectional view of a turbulent shear flow (a jet). Rotating the
figure about the $Z$ axis, the external coherent structures can be seen to be 
vortex rings. The small scale structures at the edge of the shear flow nibble at 
the ambient but do not entrain significantly.}
\label{fig:tsf}
\end{figure}
The entrainment process operates over three length scales \cite{rn}.
At the large scale, the coherent structures `engulf' parcels of ambient fluid.
The engulfed parcels then `mingle' with original shear-flow fluid 
by breaking-up/stretching into smaller scales. Subsequently, the fluids mix at
the molecular level when their characteristic sizes are comparable to 
molecular diffusion scales.

The classic example of a turbulent shear flow is a {\em jet} (see Fig. 
\ref{fig:tsf}),
which consists of a turbulent flow injected into a quiescent ambient fluid. 
Entrainment causes a jet to grow linearly in width, at a rate proportional to its 
axial velocity (Taylor's hypothesis \cite{mtt}), as a function of position along its 
direction of propagation. The width of a {\em plume} -- a hot jet -- grows linearly 
as well, but faster than a jet \cite{krs}. It is therefore deeply puzzling that 
a cumulus cloud, formed by a rising, hot, moist plume, should entrain practically nothing, and 
rise to towering heights practically undiluted. Experiments and direct numerical 
simulations on cloud-like flows \cite{bhrn,barn} show, remarkably, that it is
the local heating that is responsible for this shut-off of entrainment. This 
Letter shows that this intriguing phenomenon can be understood by studying 
the Lagrangian trajectories of fluid particles in a dynamical-systems treatment of 
a model jet-like flow, and makes further important predictions.

Our main results are as follows. We establish by quantitative methods on a model
that the entrainment and consequently the lifespan of a turbulent shear flow 
can be dramatically altered merely by tuning the relative velocities (buoyancy) 
between 
the coherent structures and the ambient. Surprisingly, a given buoyancy can either 
increase or decrease entrainment, depending on the spacing
between the structures. Secondly, we show that the {\em rate} of
engulfment is a universal function of the (approriately scaled) relative 
velocity. Although the inviscid model is incapable of estimating molecular 
mixing, we are able to make a good assessment of the mingling process
using information about the chaos in the system. The results lend themselves to
verification by direct numerical simulations and by carefully designed experiments.

I now motivate and construct the model that led to these results. The crucial
step in entrainment is the `engulfing' of ambient fluid by large-scale
coherent structures (Fig. \ref{fig:tsf}) which, as experiments as well as direct 
numerical simulations clearly show, are {\em not destroyed by heating.} It is thus
vital to understand what heating {\em does} do. Such understanding comes {\em
not} by solving for the flow behaviour in complete detail, but from an
economical model with only the essential features. To find such a model, note
that the large-scale structures responsible for entrainment in a jet can be 
approximated (Fig. \ref{fig:tsf}) by a stack of vortex rings. Vortical structures 
in the interior of the jet do not participate in the entrainment process
\cite{barn}. The mixing features of such a flow are largely
contained in an even simpler flow, that of a {\em single} pair of vortex rings. 

\begin{figure}
\epsfxsize=0.85\columnwidth
\centerline{\epsfbox{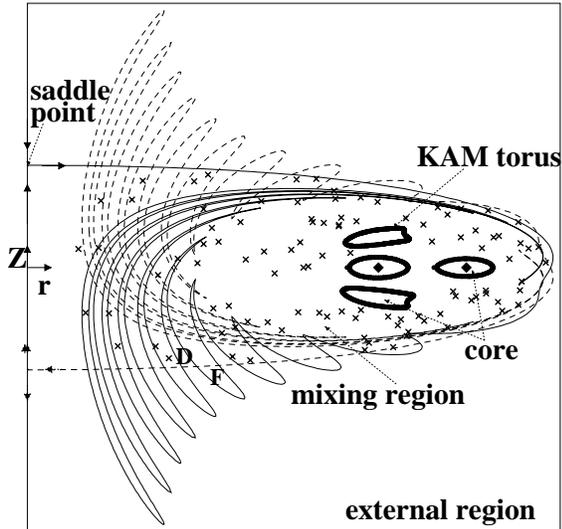}}
\caption[]{Schematic of a \P section. The unstable manifold is shown by the solid
line, the dotted line shows the stable manifold, and the crosses indicate a sample
chaotic trajectory. The phase chosen is that at 
which the two rings, indicated here by diamonds, are concentric (at $Z=0$).
The frame of reference moves with the mean axial velocity of the rings.}
\label{fig:poin}
\end{figure}
We begin with identical coaxial vortex rings of radius $R=1$ and an axial 
separation $\Delta Z$. Each ring moves forward with the sum of its 
self-induced velocity $U_s$, the induced velocity $U_i$ due to the other ring 
and the prescribed buoyancy velocity $U_b$. The rings are taken to be 
``heated/cooled'' if they are buoyant relative to the surrounding fluid, 
instantaneous acceleration is neglected and the Boussinesq approximation used. 
The fluid motion then proceeds as follows \cite{ls,glw}: 
for each vortex configuration, fluid lagrangian trajectories, obeying Hamiltonian 
dynamics, may be obtained from the Biot-Savart law. The induced velocities from the
vortices lead (a) to leap-frogging of the vortex rings, (b) to portions of the 
neighboring flow being trapped forever within KAM tori, and 
(c) to complex, often chaotic trajectories elsewhere, both within the jet and 
extending into the ambient. The dynamics in the last region is responsible for 
entrainment and subsequent mingling of ambient fluid with original jet fluid
\cite{rom,sent}. 
These features may be seen on the \P map (Fig. \ref{fig:poin}) in $r-Z$ space,
where $r$ and $Z$ are the radial and axial coordinates respectively.
The engulfment may be obtained quantitatively \cite{rom} from the tangling of 
manifolds in this map. 
The flow being axisymmetric, regions that appear as lobes in Fig. \ref{fig:poin}
are funnel-shaped structures of equal volume. During each half-cycle of leapfrog, 
the volume of fluid contained in one funnel (denoted by $F$ in the figure) is 
entrained
into the turbulent flow from the ambient, while an equivalent volume ($D$) is 
detrained from within the mixing region and dumped into the ambient \cite{rom}.
The quantity engulfed in a fraction $f$ of a cycle is equal to $2f F$.
The toroidal volume and the circulation of the rings are held respectively at 
$0.2 \pi^2$ and $4 \pi$, fixing the basic velocity and time scales. 

\begin{figure}
\epsfxsize=0.9\columnwidth
\centerline{\epsfbox{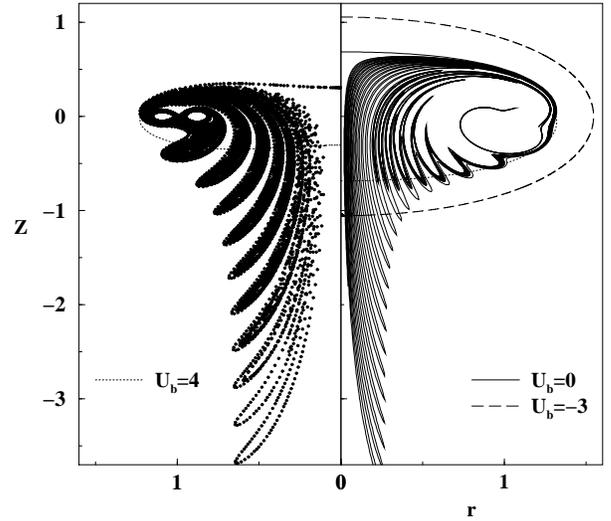}}
\caption[]{The unstable manifold for small ring spacing ($\Delta Z=0.22$). 
The stable manifold for $U_b=4$ as $t \to \infty$ is shown by the dotted line. 
(The self-induced ring velocity $U_s \sim 4.5$, so $U_b/U_s \sim 1$).}
\label{fig:m22}
\end{figure}
\begin{figure}
\epsfxsize=0.9\columnwidth
\centerline{\epsfbox{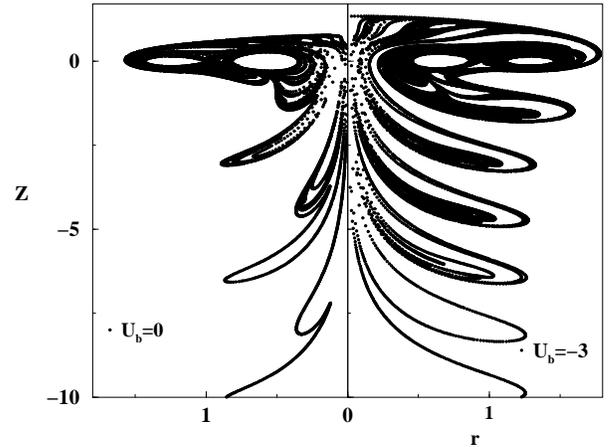}}
\caption[]{The unstable manifold for a large ring spacing ($\Delta Z = 1.00$).}
\label{fig:mone}
\end{figure}
\begin{figure}
\epsfxsize=0.9\columnwidth
\centerline{\epsfbox{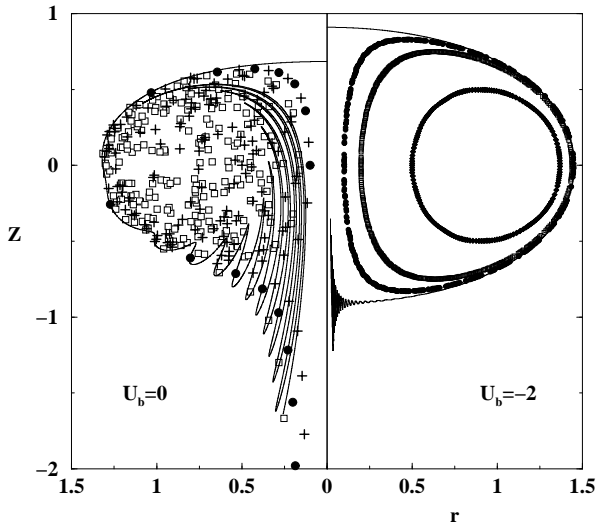}}
\caption[]{Trajectories of fluid particles when $U_b=-2$. The three
symbols represent three different starting locations.}
\label{fig:t}
\end{figure}
\begin{figure}
\epsfxsize=0.9\columnwidth
\centerline{\epsfbox{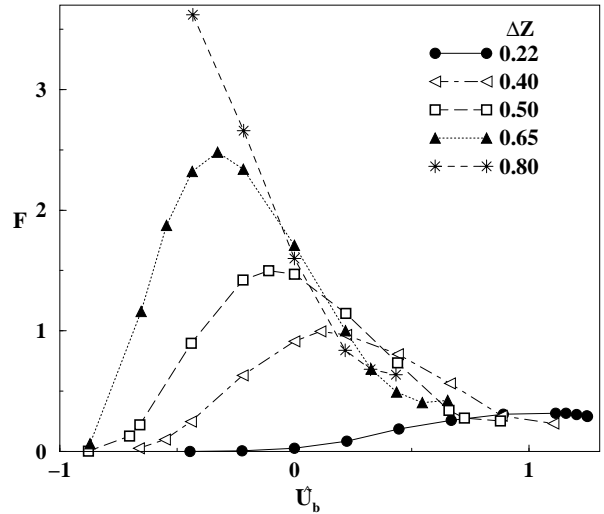}}
\caption[]{Entrainment volume as a function of buoyancy velocity imparted to
the rings. $\hat U_b = U_b/\bar U_s$ is the buoyancy velocity scaled by the average 
self-induced velocity of a ring.}
\label{fig:capF}
\end{figure}
\begin{figure}
\epsfxsize=0.9\columnwidth
\centerline{\epsfbox{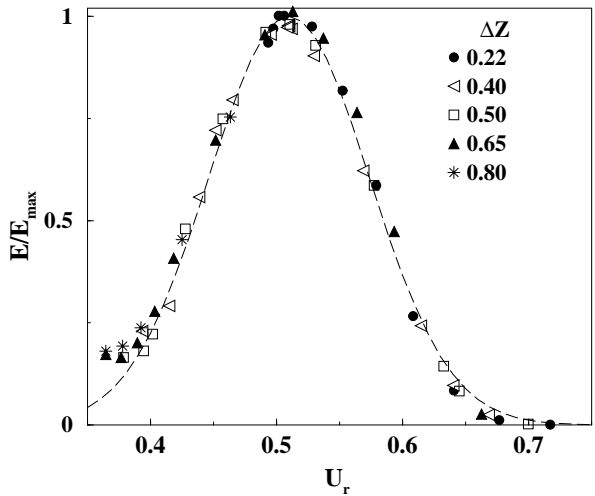}}
\caption{Dependence of the entrainment rate on the velocity ratio. The 
fit ${E /E_{\rm max}} = \exp\left\{- \left[{(U_r - {U_r}_{\rm opt}) /
\sigma} \right]^2\right\}$ is shown by the dashed line. The fitting parameters have
the numerical values $E_{max}=4.3$ and $\sigma^2 = 0.09$ here, which will change
with the circulation prescribed.}
\label{fig:fit}
\end{figure}

In the absence of buoyancy, the present predictions agree with the results of 
\cite{ptt1}, corresponding to a small $\Delta Z$, for a leap-frogging vortex pair. 
In addition, we find a maximum in the engulfment rate (proportional to the Melnikov
function \cite{wiggins}) as a function of the maximum separation $\Delta Z$ 
(inversely related to the frequency of leap-frog), which agrees with \cite{rompoje}. 
Our main results emerge when `buoyancy' is switched on: for small ring 
spacing ($\Delta Z=0.22$), the entrainment is greatly enhanced (Fig. \ref{fig:m22}) 
when the direction of buoyancy is such as to speed up the rings. In contrast, when 
the rings are slower, the entrainment is killed off almost completely. 
Surprisingly, as $\Delta Z$ increases, the role of $U_b$ is progressively 
reversed, until, at $\Delta Z = 1.0$ (Fig. \ref{fig:mone}), the engulfment is 
inversely related to $U_b$.
The extent of chaos within the region of influence
of the vortical structures provides an estimate of the `mingling' between 
original and ambient fluid. It is seen in Fig. \ref{fig:t} that for 
$\Delta Z=0.22$, when $U_b = -2$, a major portion of the region of interest 
consists of a series of imbedded KAM tori, which make any degree of mingling 
unlikely. At $U_b=-1$ (not shown) slow mixing takes place, while the 
chaotic nature of the flow at $U_b = 0$ indicates that any entering fluid will 
follow a space-filling trajectory, undergo stretching into smaller scales and 
mingle with the original fluid, resulting in a dilution of the latter. 
Fig. \ref{fig:capF} shows the dependence of $F$ on the buoyancy for different
values of ring spacing. For the highest spacing, due to the extreme tangling of
the manifolds, the computations were reliable only for the range of $U_b$ shown.
When $\Delta Z$ is about half the radius, the 
engulfment decreases significantly {\em whether the fluid is heated or cooled}. 
For each ring spacing
the engulfment displays a maximum at some value of the buoyancy velocity $U_b$. 
Table 1 shows characteristic axial velocities for different $\Delta Z$.
Here $\hat U_i$ is the maximum axial velocity, scaled by $\bar U_s$, induced by 
one of the rings on the other during a given cycle, ($\hat U_i$ would
also be the typical velocity induced by one of the rings on a non-vortical 
fluid element at the same distance.) $[\hat U_b]_{\rm opt}$ is the buoyancy 
velocity required to maximize engulfment at a given $\Delta Z$.
The last column in the table is independent of $\Delta Z$, which suggests that 
we define a velocity ratio $U_r$:
\be
U_r \equiv {U_i \over U_i + U_b + c \bar U_s},
\label{krat}
\ee
whose numerator is a measure of the velocity of the
induced motion in the surrounding fluid, and denominator represents the
total ring speed. The constant $c$ ($1.55$ here) depends on the ratio of the 
ring thickness to the toroidal radius, as well as on the particular measures of 
the self and mutually induced velocities used. 
Since the period $\tau$ of a leapfrog is vastly different for different 
ring spacings, the {\em rate} of engulfment $E \equiv F / \tau$,  would be a 
more appropriate measure for comparison. This quantity is plotted in Fig. 
\ref{fig:fit} as a function of the velocity ratio $U_r$.

\begin{center}
{\bf Table 1}
\vskip 0.05in
\begin{tabular}{|c|c|c|c|c|}
\hline
&&&& \\
$\ \Delta Z \ $ &   $\hat U_i$     & $[\hat U_b]_{\rm opt}$    & 
 ${\bar{U}}_s$ & $\hat U_i-[\hat U_b]_{\rm opt}$  \\
&&&& \\
\  0.22 \  & \   2.676 \ &\   1.130 \ & \  4.493 \ &  1.55   \\
\  0.40 \  & \   1.689 \ &\   0.128 \ & \  4.531 \ &  1.56   \\
\  0.50 \  & \   1.450 \ &\  -0.110 \ & \  4.556 \ &  1.56   \\
\  0.65 \  & \   1.236 \ &\  -0.327 \ & \  4.591 \ &  1.56   \\
\  0.80 \  & \   1.109 \ &\  -0.433 \ & \  4.620 \ &  1.54   \\
\hline
\end{tabular}
\end{center}

The rate of entrainment is seen to be a universal function of the velocity ratio,
i.e., the parameter which determines the entrainment rate is the ratio of the motion
induced in the surroundings to the speed of the coherent structures.
It is observed that the entrainment when the rings are 
much faster than the surroundings (extreme left of Fig. \ref{fig:fit}) levels off 
at a small value. This feature will be investigated in detail elsewhere.

In the simulations of \cite{barn}, the structure spacing is reduced upon heating 
from about $\Delta Z=1$ to $\Delta Z=0.3$. The ratio $U_r$ in the present 
scales is about $0.45$ in the pre-heating zone, where we expect a significant level 
of entrainment; and about $0.7$ in the heating zone, where entrainment should drop 
almost to zero; which concur with what the simulations show.
In a lab jet (or in a cloud), the entire turbulent flow would be heated with
respect to the ambient, not the structures alone. Moreover, the rings would be of
unequal sizes and strengths. We have examined these situations: when the entire 
internal fluid, or a selected portion of it, is given the same buoyancy as the 
rings, the results are very close to those shown above, demonstrating
that the important parameter is the relative velocity between the structures and 
the {\em external} flow, and not the state of buoyancy of the interior fluid.
Inequalities in the rings result in one ring becoming dominant over the other, but
have very little qualitative effect on the results. 

In summary, it is known from observations that the entrainment in heated plumes 
is higher than in unheated jets, while that in (heated) clouds is much lower.
This is the first computation to our knowledge that addresses this apparent
contradiction, obtains entrainment quantitatively albeit for a model heated flow, 
and shows that an imposed relative buoyancy can {\em either increase or decrease the 
entrainment}. For example in Fig. \ref{fig:fit}, when $\Delta Z=0.5$, the maximum 
entrainment occurs when $U_b = 0$, speeding-up or slowing-down the rings would both
decrease entrainment. For small values of structure spacing, the 
entrainment is directly related to the buoyancy (with the sign as defined here), 
while while at large spacing they are inversely related. 
The data collapse seen in Fig. \ref{fig:fit} could not have been anticipated from 
intuition, and much less from an examination of the tangling of the manifolds. 
The funnel shapes and the convolutions they undergo are often qualitatively 
different in each case (they sometimes even have two or three sub-lobes 
per half-cycle), and so is the volume they entrain, but the 
rate of entrainment obeys a very simple relationship. Implicit in the data
collapse is the remarkable observation that the maximum rate of engulfment (for 
a given circulation and geometry) is independent of the system frequency. 
An increase in the engulfment of external fluid due to buoyancy is shown to be 
accompanied by an enlargement of the chaotic region, resulting in increased
mingling. In closing, note that we have neglected
viscous effects as well as the complicated manner in which heating/cooling is bound
to affect a turbulent shear flow, by altering the geometry, density and orientation 
of the coherent structures, which would in turn affect entrainment. Including these
would be natural directions for future research. In order to test the present 
predictions and to strengthen our understanding the effect of heat,
systematic experiments and direct numerical simulations are called for. 
The ideal experiment would be on a
laboratory-generated isolated pair of vortex rings, whose circulations could be
estimated and which could be heated or cooled over some axial distance.

I thank A. Leonard, S. Wiggins, R. Narasimha and K. R. Sreenivas for 
discussions and useful suggestions. This work is funded by the AR\&DB, India.

\end{document}